\documentclass{article}  \usepackage{epsfig} \usepackage{float}

\begin{document}

\title{Improved study of a possible  $\Theta^+$ production in the
$pp\rightarrow{}pK^0\Sigma^+$ reaction with the COSY-TOF spectrometer}
\author
{
{\bf The COSY-TOF Collaboration}\\ 
M.~Abdel-Bary$^3$,
S.~Abdel-Samad$^3$,
K.-Th.~Brinkmann$^1$,
R. Castelijns$^3$,\\ 
H.~Clement$^4$,
J.~Dietrich$^1$,
S.~Dshemuchadse $^1$,
E.~Dorochkevitch $^4$,
W.~Eyrich$^2$ \footnote{corresponding author, email:wolfgang.eyrich@physik.uni-erlangen.de},\\ 
K.~Ehrhardt$^4$,
A.~Erhardt$^4$,
H.~Freiesleben$^1$,
W.~Gast$^3$,
J.~Georgi$^2$,\\ 
A.~Gillitzer$^3$,
H.~J\"ager$^3$,
R.~J\"akel$^1$,
L.~Karsch$^1$, 
K.~Kilian$^3$,
M.~Krapp$^2$,
E.~Kuhlmann$^1$,\\
A.~Lehmann$^2$,
H.~P.~Morsch$^3$,
N.~Paul$^3$, 
L.~Pinna$^2$,
C.~Pizzolotto$^2$,
J.~Ritman$^3$\\
E.~Roderburg$^3$,
S.~Schadmand$^3$,
P.~Sch\"onmeier$^2$,
M.~Schulte-Wissermann$^1$,\\ 
W.~Schroeder$^2$,
T.~Sefzick$^3$,
A.~Teufel$^2$,
A.~Ucar$^3$,\\ 
W.~Ullrich$^1$,
R.~Wenzel$^1$,
P.~Wintz$^3$,
P.~W\"ustner$^3$,
P.~Zupranski$^5$  \\ 
  $^1 $ {\small Institut f\"ur Kern- und Teilchenphysik, Technische Universit\"at Dresden, D-01062 Dresden}\\
  $^2 $ {\small Physikalisches Institut, Universit\"at Erlangen-N\"urnberg, D-91058 Erlangen}\\
  $^3 $ {\small Institut f\"ur Kernphysik, Forschungszentrum J\"ulich, D-52425 J\"ulich}\\
  $^4 $ {\small Physikalisches Institut, Universit\"at T\"ubingen, D-72076 T\"ubingen}\\
  $^5 $ {\small Andrzej Soltan Institute for Nuclear Studies, PL-00681 Warsaw }\\
}

\date{\today} 
\maketitle
\begin{abstract}
The $pp\rightarrow{}pK^0\Sigma^+$ reaction was investigated with the TOF
spectrometer at COSY at $3.059~{\rm{}GeV/c}$ incident beam momentum.
The main objective was to clarify whether or not a narrow exotic
$S=+1$ resonance, the $\Theta^+$ pentaquark, is populated at
$1.53~{\rm{}GeV/c}^2$ in the $pK^0$ subsystem with a data sample of
 much higher statistical significance compared to the previously
 reported data in this channel.
An analysis of these data does not confirm the existence of the $\Theta^+$
pentaquark. This is expressed as an upper limit for the cross section
$\sigma(pp\rightarrow{}\Sigma^+\Theta^+)<0.15~{\rm{} \mu b}$ at the 95\% confidence level.
\end{abstract}

\section{Introduction}

The quark model introduced by Gell-Mann in 1964~\cite{MGell-Mann64} successfully
explains strongly interacting particles, observable as being
colorless systems of either a quark-antiquark pair for mesons or of
three quarks for baryons.
However, QCD does not exclude the existence of other color singlet
objects containing additional quark-antiquark pairs or gluons as constituents.
Within a chiral soliton model Diakonov, Petrov, and Polyakov~\cite{DDiakonov97}
predicted the existence of an anti-decuplet of baryonic states with $J^p=1/2^+$
consisting of four quarks and one anti-quark.
Three members of this anti-decuplet are manifestly exotic, having 
combinations of strangeness and isospin not allowed for three-quark systems.
The lightest of these exotic states is the $\Theta^+$ pentaquark with a quark
content of $uudd\bar{s}$ and thus strangeness $S=+1$,  a predicted mass
of about
1.53~GeV/c$^2$~\cite{DDiakonov97}, and rather narrow width of less than
15~MeV~\cite{MPolyakov00}.

The first experimental evidence for a narrow $S=+1$ baryonic resonance in the
predicted mass region came from the LEPS collaboration at
SPring8~\cite{TNakano03} who observed a peak in the $K^+n$ invariant system at
1.54~GeV/c$^2$ in the $\gamma{}n\rightarrow{}K^+K^-n$ reaction on $^{12}$C.
Shortly after, the DIANA collaboration found evidence in the $K^0p$ system in a
re-analysis of bubble chamber data of the reaction
$K^+{\rm{}Xe}\rightarrow{}K^0pX$~\cite{VVBarmin03}.
Soon after these first observations positive signals indicating the existence of
a narrow resonance in this mass region were also reported by
CLAS~\cite{SStepanyan03,VKubarovsky04}, SAPHIR~\cite{JBarth03},
HERMES~\cite{AAirapetian04}, ZEUS~\cite{SChekanov04},
COSY-TOF~\cite{MAbdel-Bary04}, SVD-2~\cite{AAleev04}, by Asratyan
{\it et al.}~\cite{AEAsratyan03}, and by Aslanyan
{\it et al.}~\cite{PZhAslanyan04} in a large variety of reactions including
even neutrino-induced processes~\cite{AEAsratyan03}.

Evidence from these experiments was claimed for the
existence of a narrow state in the systems $K^0p$ and $K^+n$ in the mass region
between 1.522 and 1.555 GeV/c$^2$.
In each of these cases the signal is based on at most 50
events. Thus, none of these experiments has the statistical accuracy required
to establish the existence of the $\Theta^+$.
Moreover, several experiments particularly at high energies including
BES~\cite{JZBai04}, BABAR~\cite{BAubert05}, Belle~\cite{KAbe04},
HERA-B~\cite{KTKnoepfle04}, SPHINX~\cite{YuMAntipov04},
HyperCP~\cite{MJLongo04}, CDF~\cite{DOLitvintsev05}, and FOCUS~\cite{KStenson05}
did not confirm the state and set upper limits on the production
cross section.
Although the production mechanism of the $\Theta^+$ pentaquark in the
high energy reactions studied may differ strongly from that for conventional baryonic
states, the null results cast additional doubt on the existence of the
$\Theta^+$.

In the meantime a series of improved experiments and analyses were carried out
by the groups who had reported a positive signal in their first experiments
in order to obtain data with higher statistical accuracy.
Recently, the results of dedicated experiments at CLAS were published.
These yielded negative results both on hydrogen~\cite{MBattaglieri06} and
deuterium targets~\cite{SNiccolai06}.
The high statistics null result on the hydrogen target is in clear contradiction
to the SAPHIR result~\cite{JBarth03}.
The negative result on the deuterium target does not agree with the earlier
positive result of CLAS in the channel $\gamma{}d\rightarrow{}pK^-\Theta^+$ and
sets an upper limit to the $\Theta^+$ production cross section.
In contrast, SVD-2~\cite{AAleev05} and DIANA~\cite{VVBarmin06}
analyzed additional data which were not included in the first publications with
optimized analysis methods.
Both groups confirmed their previous positive results with increased significance.
Furthermore, DIANA estimated the
$\Theta^+$ width to be only 0.4~MeV.
A recent measurement at KEK~\cite{KMiwa06} of the channel
$\pi^-p\rightarrow{}K^-X$ 
revealed a bump at 1.530 GeV/c$^2$ in the
$K^-$ missing mass spectrum with a significance of  2.5~$\sigma$.
A recent dedicated experiment of the LEPS Collaboration
shows a peak around 1.530~GeV/c$^2$ in the channel
$\gamma{}d\rightarrow{}pK^-\Theta^+$~\cite{TNakano06}, for which CLAS did not
confirm its previous positive result.
However, the detector acceptance regions covered by these experiments
are different.

Recent theoretical studies by  Sibirtsev {\it et al.}~\cite{Sibirtsev-04}\cite{Sibirtsev-05}, 
by Roberts~\cite{Roberts}, and by Diakonov~\cite{DDiakonov06}
obtain small values of about 1~MeV as upper limit for the $\Theta^+$ width. Ref.~\cite{DDiakonov06}
reviews the present experimental status and concludes that for a width below 1~MeV as estimated 
by DIANA~\cite{VVBarmin06} the existence of the $\Theta^+$ is not ruled out by the recent negative 
results including those of CLAS \cite{MBattaglieri06}\cite{SNiccolai06}. In this situation further 
measurements with improved accuracy using different reactions were needed to clarify the $\Theta^+$ puzzle. 

In this letter we report on the results of an experiment studying the
$pp\rightarrow{}pK^0\Sigma^+$ reaction with the COSY-TOF spectrometer with
substantially improved statistical accuracy and extended detection
capability.
COSY-TOF is the only experiment involved in $\Theta^+$ studies which
provides an exclusive
measurement of the final state in $pp$ collisions, and it is unique in its almost
complete coverage of the three-body phase space.
Moreover the exclusive measurement selects the strangeness in the 
$K^0p$ system to be $S=+1$, which is a unique feature for experiments searching for the 
$\Theta^+$ in the $K^0p$ system.

In the following the improved experimental setup will be described and
the results obtained by three independent analyses will be presented. 

\section{Experiment}
The experiment was performed at the cooler synchrotron COSY with the COSY-TOF
spectrometer \cite{web}. The preceeding measurement \cite{MAbdel-Bary04} was
carried out at a beam momentum of 2.95 GeV/c, a slightly higher momentum of
3.059 GeV/c was chosen for the new measurement. At this higher momentum the
upper bound of the  $K^0p$ invariant mass is  1.597 GeV/c$^2$ and therefore a
possible structure at 1.530 GeV/c$^2$ is further removed from the upper mass
limit.

\begin{figure} [H] 
\begin{center}
  \epsfig{file=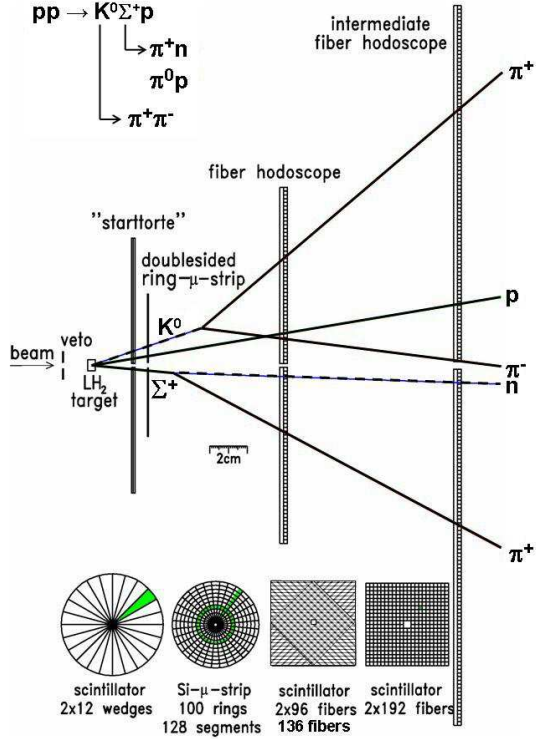,width= 8 cm}
\caption{\label{startcounter}Schematic view of the start region with the track
pattern of a  $pp \rightarrow pK^0\Sigma^+$ event}
\end{center}
\end{figure}

The reaction $pp \rightarrow pK^0\Sigma^+$ is induced by focusing the  proton
beam on a spot of about 1 mm in diameter on a liquid hydrogen target (length 4
mm, diameter 6 mm).  The start detector system, which was upgraded for the
experiment described here, is mounted 2 cm downstream of the target.
A schematic view is shown in Fig. \ref{startcounter} together with the event
topology of the studied reaction.  The first detector is a thin scintillation
counter with a diameter of 15 cm, consisting  of two segmented layers of 1~mm
thickness. Each layer is subdivided into 12 wedge shaped scintillators.  This
detector provides the multiplicity of the charged particles close to the
target as well as  the start signal for the time-of-flight measurement. It is
followed by a double-sided silicon  microstrip detector with a diameter of
about 6 cm, the front and rear sides of which  consist of 100 rings and  128
sectors, respectively, giving a precise track point close to the target.  At a
distance of about 10 cm and 20 cm two scintillating fiber hodoscopes are
installed to obtain two further track points.  These detectors are required to
identify the tracks from the delayed decay of $K^0_S$ mesons into
$\pi^+\pi^-$ pairs. The track of the $K^0$- meson is defined by the vertex of
its decay and the primary vertex.  For the measurement described here the
first hodoscope was replaced  by an upgraded version.  It consists of two
crossed planes and a stereo plane at $45^0$, all built of squared fibers
(2~mm $\times$ 2~mm), forming an inner part of about 20~cm $\times$ 20~cm with
three layers and an outer part of about twice the size with two layers. The
second hodoscope with a size of about 40~cm$\times$ 40~cm consists of two
crossed layers. The inner detector system covers the full phase-space of the
primary particles ($\Sigma^+$-hyperon, $K^0$-meson, and proton) apart from
tiny holes with a diameter of about 2 mm for the beam.

This also holds for the outer detector system, which consists of 96
scintillating bars arranged as a barrel with radius of 1.5 m and length of 2.8
m, and a forward wall at 3.3 m distance from the target
\cite{quirl}\cite{barrel}.  The outer detector provides the stop signal for
the time-of-flight measurement as well as  additional track points.

The trigger condition is based on the increase in charged-particle
 multiplicity between the start scintillator and the outer detector system,
 which results from the $K^0_S \rightarrow \pi^+\pi^-$ decay (see
 Fig.\,\ref{startcounter}). The used trigger reduces the event rate by a factor of about $10^4$.

The TOF spectrometer supplies very good position information due to its high
granularity allowing track reconstruction. Together with time-of-flight
information this is exploited in the three analyses of the data in various
ways, as described in the following section.

\section{Data Analysis}

The analysis was carried out with three independent analysis programs (A, B,
C), which differ in algorithms and event selection methods, but are based on a
common calibration of all detector components. In order to cross-check each
analysis the complete data sample was divided into two independent parts
(events recorded with either even or odd number). The analysis procedures were
developed using even events only and the odd events were analyzed only after
the analysis procedures were frozen.
 
 A $pK^0\Sigma^+$ event is identified according to Figure \ref{startcounter}
 by its topology, that is a prompt track emerging from the target (proton), a
 delayed decay ($K^0$) and optionally a kink in a charged track ($\Sigma^+$).

 A prompt track emerging from the target is assigned to a proton. The $\chi^2$
of a straight line fit  and the number of detector elements with hits on this
track are used as quality criteria,  which serve for the track selection. The
selection parameters are different in the three analyses.  The momentum is
calculated from the time-of-flight and the proton mass. The $K^0_S$ is
reconstructed  from the invariant mass of the decay particles
$\pi^+\pi^-$. The secondary vertex is reconstructed  from the fitted pion
tracks and restricted to the region between the microstrip detector and the
first fiber hodoscope. The plane defined by these tracks must contain the
primary vertex. In order to  discriminate against the $\Lambda \rightarrow
p\pi^-$ decay the different analyses place different  selections on the angles
of the decay particles with respect to the $K^0_S$ direction and on their
velocities. The $\Sigma^+$ is identified by a kink in a charged particle
track. The direction of the  $\Sigma^+$ is measured with the silicon micro
strip detector. 

The main aspects in which the analyses differ, are the 
following: AnalysisA: The $\Sigma^+$ is reconstructed from the missing mass of 
the p$K^0$system, the value of the missing mass is used as the final selection
criterion.  As the $\Sigma^+$ signal in the microstrip detector is not 
required, no constraint  on the $\Sigma^+$ decay vertex is imposed. 
Analysis B:  The detection of a kink due to the $\Sigma^+$ decay  is
required.  The direction of the $\Sigma^+$ is calculated from the microstrip
detector pixel and the primary vertex. For the two subsystems $\pi^+\pi^-$ and
$p\Sigma^+$ the invariant and missing mass are calculated, respectively.
Both calculations require the direction of $K^0$ in addition. If there is
more than one candidate topology for a $pK^0\Sigma^+$ event, the topology
with  the minimum difference of both calculated masses to the  $K^0$ mass is
chosen.   A 3C-kinematic fit is applied to the final event. Event selection is
based on the  subsystem masses and on the $\chi^2$ of the kinematic fit.
Analysis C:  The $\Sigma^+$ is identified as in analysis B.  A 3C-kinematic
fit is applied for all possible topologies of one event, that with the lowest
$\chi^2$  is selected, if the $\chi^2$ is lower than a given limit. 

In all analyses an instrumental background from incorrectly reconstructed
events in  the data sample was investigated by various methods. These include
side band  investigations and variations of the maximum $\chi^2$. The
background was found to  be smooth in the $pK^0$, $p\Sigma^+$ and
$K^0\Sigma^+$ invariant mass distributions.  The background is determined to
be 21\% (A), 25\% (B), 28\% (C). 

The corrections for detector acceptance and 
for the efficiencies of the analyses are  deduced from Monte Carlo simulations 
that use an equal population of 3-body phase space as input. The efficiencies 
of the analyses (including all branching ratios) are  
$(2.0\pm 0.1)\cdot 10^{-3}$ (B,C) and $(4.2\pm 0.2)\cdot 10^{-3}$ (A). 
The latter one  is larger since there is no restriction on the location of the $\Sigma^+$-decay vertex. 
The resolution in the invariant mass distribution 
of the $pK^0$ subsystem is $\sigma = 6~\rm{}MeV/c^2$ for analyses A,C 
and $5~\rm{}MeV/c^2$ for analysis B as deduced  from Monte Carlo.

Elastically scattered events were recorded in parallel in order to determine
the luminosity. In total $1.3\cdot 10^{10}$ events were recorded with an
integrated luminosity of $(214\pm 20)~\rm{}nb^{-1}$. Approximately 4000 
$pK^0\Sigma^+$ events in analysis  B,C, and 7900 events in analysis A, 
are reconstructed. Due to the different strategies  of the analysis about 
300 events are found both in B and C. For each combination A,B  and A,C about 
600 events are shared. Therefore, in total more than 12.000 independent  
$pK^0\Sigma^+$ events were reconstructed.


\section{Evaluation of results}

The total cross section, obtained by the different analyses, is given in Table
1. Only statistical errors are quoted. The systematic error of the total
cross section is estimated to be $\pm$ 1 $\mu$b. It is mainly caused by
uncertainties in the determination of both the luminosity and the detection
and reconstruction efficiencies.

\begin{figure}[H]
\begin{center}
  \epsfig{file=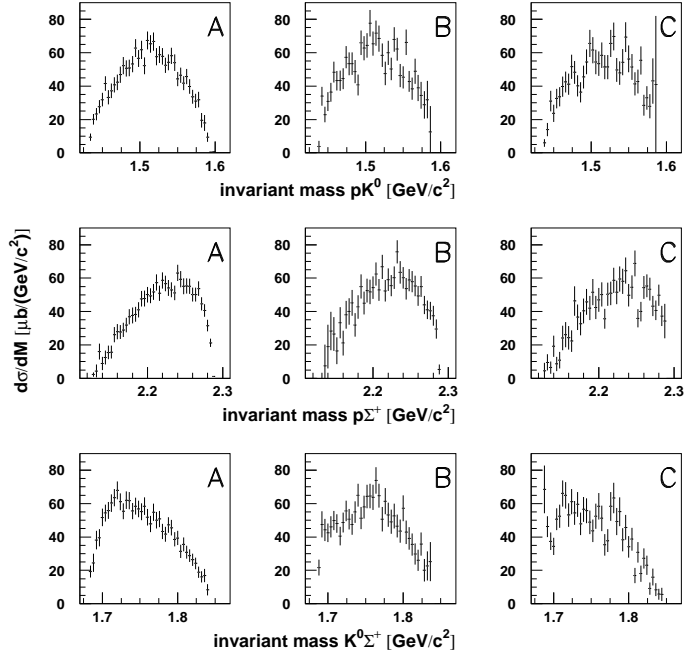,width=10cm}
\caption{\label{invMass1} The invariant masses for the three subsystems in the
  $pK^0\Sigma^+$ final state. The first column shows the results of analysis
  A, the second one that of analysis B, the third one that of analysis C. }
\end{center}
\end{figure}

\begin{figure}[H] 
\begin{center}
  \epsfig{file=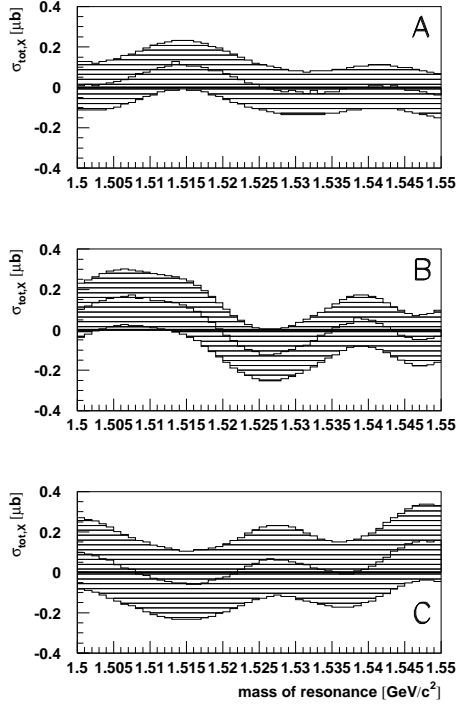,width=7cm}
\caption{\label{conflev} 95\% confidence range for the cross section of a narrow resonance as a function of 
  $M_{pK^0}$ for the A, B, C analyses in the top, center, and lower frames,
  respectively. The central lines of  each band present the contribution of a 
  hypothetical narrow resonance with the lowest $\chi^2$ value.}
\end{center}
\end{figure}

\begin{table}[H] 
\begin{center}
\begin{tabular}{c|c}
 analysis & cross section\\  \hline A &6.9 $\mu$b $\pm$ 0.2 $\mu$b\\ B
 &7.0 $\mu$b $\pm$
 0.3 $\mu$b\\ C &7.0 $\mu$b $\pm$ 0.3 $\mu$b\\
\end{tabular}
\end{center}
\caption{Total cross section of $pp\rightarrow{}pK^0\Sigma^+$ }
\end{table}
The invariant mass spectra of the subsystems $pK^0$, $p\Sigma^+$, and
$K^0\Sigma^+$  as extracted from the three analyses are shown in
Fig. \ref{invMass1} after background subtraction and acceptance correction.
All results agree within statistical uncertainties. No significant differences of 
spectra obtained from even and odd events were found. The invariant mass spectra do not follow
an equal population of phase space. However this is without implication for the search for the 
$\Theta^+$.
\begin{figure}[H]
\begin{center}
  \epsfig{file=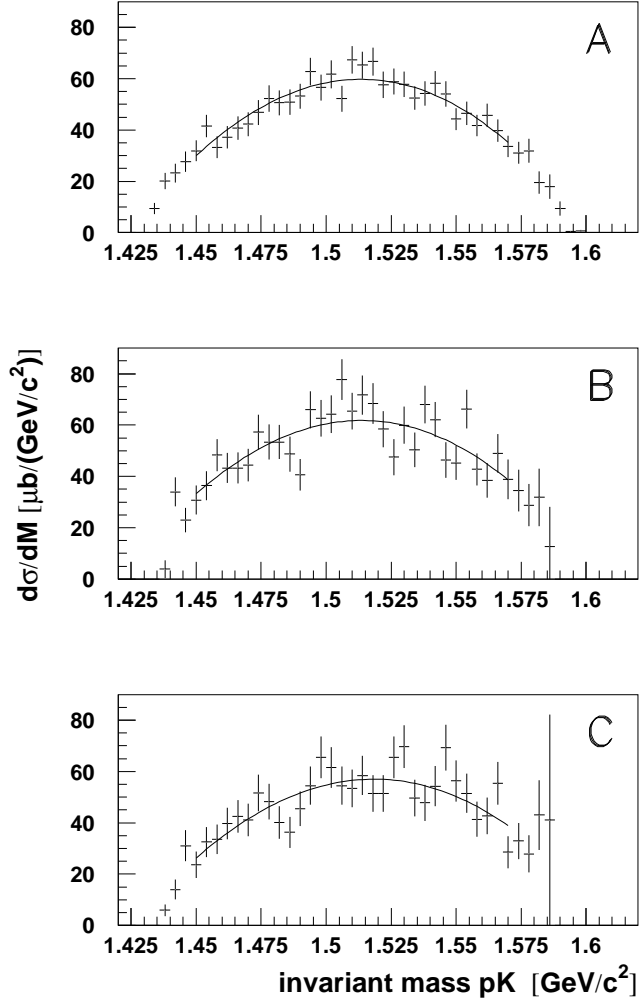,width=10cm}
\caption{\label{invMass2} The invariant masses of the 
  $pK^0$ system for the  three analyses together with a 3$^{rd}$ order polynomial parameterization.}
\end{center}
\end{figure}
The $pK^0$ mass spectra are presented again in Fig.\,\ref{invMass2} together with a 
$3^{rd}$ order polynomial parameterization in the mass region of 
$1.45~\rm{}GeV/c^2 < M_{pK^0} < 1.57~\rm{}GeV/c^2$. They were analyzed in order to determine the statistical
significance with which a narrow structure might be present.
A narrow structure was added to the polynomial described above.
The shape of this narrow structure has been taken from Monte
Carlo simulations   of a resonance with a width negligible
compared to  the detector resolution. The width of the added structure therefore corresponds to the 
experimental resolution of the invariant $pK^0$ mass of $\sigma$=~$6\,MeV/c^2$ for analyses A,C and 
$\sigma$=~$5\,MeV/c^2$ for analysis B (see chapter 'data analysis')
The mass of the resonance was varied in $1~\rm{}MeV/c^2$ steps over the
$M_{pK^0}$ range from
1.50~$\rm{}GeV/c^2$ - 1.55~ $\rm{}GeV/c^2$. 
The strength of the structure for each setting was varied between  $-1~\mu\rm{}b < \sigma_{tot,X} < +1~\mu\rm{}b$.
The negative values here correspond to a drop of the $M_{pK^0}$ differential cross section
below the expectation based on the polynomial parameterization. 
For each combination  of $M_{pK^0}$  and $\sigma_{tot,X}$ the value of $\chi^2$ of the fit to the data
was determined. These results are summarized in Figure \ref{conflev}.
   In this figure the value of $\sigma_{tot,X}$ corresponding to
the minimal value of $\chi^2$ 
is represented by the central line of the band  as a function of $M_{pK^0}$.
The 95\% confidence interval for an enhancement or suppression of the measured 
$M_{pK^0}$ differential cross section is indicated by the width of the band. 
The results presented in Figure \ref{conflev} indicate that over the full
$M_{pK^0}$ range investigated here the parameterization assuming
$\sigma_{tot,X}$ = 0~$\mu\rm{}b$ is consistent with the measured data within the
95\% confidence level. In particular, this new, higher statistics data
do not contain positive evidence for a narrow
structure at $M_{pK^0}$ =1.530~$\rm{}GeV/c^2$.
The fluctuation of the central value of the 95\% confidence intervals are  not
correlated between the different analyses.
Based upon the smallest upper limit of the three  95\% confidence intervals the 
maximum cross section for a narrow resonance   $\sigma_{tot,X} < $ 0.15~$\mu\rm{}b$ has
been deduced over the full mass range.

These results also have to be compared with the positive evidence for a
$\Theta^+$ resonance found in the data of the first measurement \cite{MAbdel-Bary04}.
 First of all the number of reconstructed events is much
larger for the new measurement - about a
factor of 4 for analyses B and C and about a
factor of 8 for analysis A - which gives a
strongly improved statistical accuracy. In addition, the three largely different
analyses give a better understanding of systematic uncertainties. 
In the previous measurement the peak was located 32 MeV/c$^2$  
below the upper kinematic limit of the populated $pK^0$ mass range
and therefore resided on a continuum background with a steep negative
slope. This caused uncertainties in the determination of the background.
Therefore, in the present experiment a higher excess energy was chosen. 
Comparing the $pK^0$ spectra obtained in this work with the published one \cite{MAbdel-Bary04}
a significant difference in the shape, which is more symmetric in the new
measurement, is noticed.
Since the $pK^0$ invariant mass distribution obtained in the new experiment is based on three
independent analysis procedures including systematic studies of the instrumental background, we
are confident in the symmetric shape of the  $pK^0$  continuum. 
The data points of the three different analyses together with their uncertainties define a band, 
which represents the continuum shape of these analyses.
\begin{figure}[H] 
\begin{center}
\epsfig{file=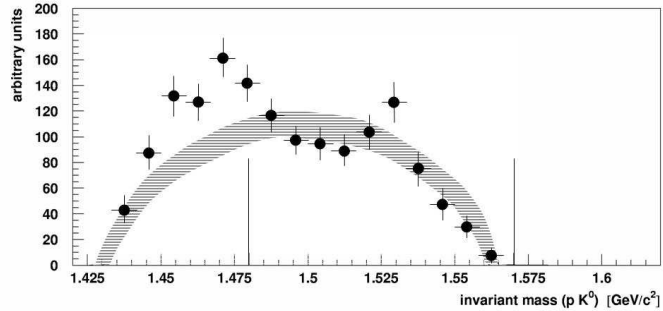,width=9cm}
\caption{\label{olddata} Invariant mass of the $pK^0$ spectrum of the previous measurement together with a 
band representing the shape of the new measurement. The height of the band is adjusted in the mass range 
indicated by the two vertical lines.}
\end{center}
\end{figure}
In order to compare it to the previous data, this band is scaled to the kinematically allowed range.
Fitting its height within a mass region between 1.48 GeV/c$^2$ and 1.57 GeV/c$^2$, symmetric with 
respect to the peak observed at 1.53 GeV/c$^2$ \cite{MAbdel-Bary04}, the significance of the peak decreases 
substantially if this shape is used as continuum for the previous data (see Figure~\ref{olddata}).
\section{Summary}

The reaction $pp\rightarrow{}pK^0\Sigma^+$ was studied in an exclusive
measurement at a beam momentum of 3.059 GeV/c with complete phase space
coverage. The extracted $pK^0$ spectra do not show  evidence for a narrow
resonance  in the mass region of 1.50 GeV/c$^2$ - 1.55 GeV/c$^2$ in any of the three
independent analyses. The data are consistent with a cross section of
$\sigma_{tot,X}=0\,\mu$b and an upper limit of 0.15 $\mu$b is derived with a
confidence level of 95\%. The evidence for a $\Theta^+$, which was found in a
preceding measurement, is not confirmed.
\section*{Acknowledgment}
We would like to thank very much the COSY accelerator team for the preparation
of the excellent proton beam and for the good cooperation. Moreover we are
grateful to J. Sarkadi and J. Uehlemann  for technical support.  This
experiment is supported by the German BMBF and FZ J\"ulich.

\end{document}